\pdfoutput=1
\documentclass[12pt,a4paper]{article}

\usepackage{ifthen} 
\newboolean{pdflatex}
\setboolean{pdflatex}{true} 
\newboolean{articletitles}
\setboolean{articletitles}{true} 

\newboolean{uprightparticles}
\setboolean{uprightparticles}{false} 

\def\paperauthors{LHCb collaboration}
\def\paperasciititle{Measurement of CP asymmetry in D0 -> KS KS decays with with Run 3 data} 
\def\papertitle{Measurement of $C\!P$ asymmetry in $D^0 \to K^0_{\rm S} K^0_{\rm S}$ decays with Run~3 data} 
\def\paperkeywords{{High Energy Physics}, {LHCb}} 
\def\papercopyright{\the\year\ CERN for the benefit of the LHCb collaboration} 
\def\paperlicence{CC BY 4.0 licence}
\def\paperlicenceurl{https://creativecommons.org/licenses/by/4.0/}

\newif\ifEnableSectionTOCLinks
\EnableSectionTOCLinksfalse

\usepackage[top=1in, bottom=1.25in, left=1in, right=1in]{geometry}

\usepackage{microtype}
\usepackage{lineno}  
\usepackage{xspace} 
\usepackage{caption} 

\usepackage{graphicx}  
\usepackage{color}
\usepackage{colortbl}
\graphicspath{{./figs/}} 

\usepackage{amsmath} 
\usepackage{amssymb}
\usepackage{amsfonts}
\usepackage{upgreek} 

\newcommand*\patchAmsMathEnvironmentForLineno[1]{%
\expandafter\let\csname old#1\expandafter\endcsname\csname #1\endcsname
\expandafter\let\csname oldend#1\expandafter\endcsname\csname
end#1\endcsname
 \renewenvironment{#1}%
   {\linenomath\csname old#1\endcsname}%
   {\csname oldend#1\endcsname\endlinenomath}%
}
\newcommand*\patchBothAmsMathEnvironmentsForLineno[1]{%
  \patchAmsMathEnvironmentForLineno{#1}%
  \patchAmsMathEnvironmentForLineno{#1*}%
}
\AtBeginDocument{%
\patchBothAmsMathEnvironmentsForLineno{equation}%
\patchBothAmsMathEnvironmentsForLineno{align}%
\patchBothAmsMathEnvironmentsForLineno{flalign}%
\patchBothAmsMathEnvironmentsForLineno{alignat}%
\patchBothAmsMathEnvironmentsForLineno{gather}%
\patchBothAmsMathEnvironmentsForLineno{multline}%
\patchBothAmsMathEnvironmentsForLineno{eqnarray}%
}

\usepackage[pdftex,
            pdfauthor={\paperauthors},
            pdftitle={\paperasciititle},
            pdfkeywords={\paperkeywords}]{hyperref}
\usepackage{hyperxmp}
\hypersetup{
    pdfcopyright={Copyright (C) \papercopyright},
    pdflicenseurl={\paperlicenceurl}
}

\usepackage[colorinlistoftodos,textsize=scriptsize]{todonotes}

\usepackage[bottom,flushmargin,hang,multiple]{footmisc}

\usepackage[all]{hypcap} 

\usepackage{xspace} 
\usepackage{upgreek}

\def\lhcb   {\mbox{LHCb}\xspace}

\def\MagUp {\mbox{\em Mag\kern -0.05em Up}\xspace}
\def\MagDown {\mbox{\em MagDown}\xspace}

\def\hltone {HLT1\xspace}
\def\hlttwo {HLT2\xspace}

\ifthenelse{\boolean{uprightparticles}}%
{

 \def\Ppi         {\ensuremath{\uppi}\xspace}

 \def\PDelta      {\ensuremath{\Delta}\xspace}                 
 \def\PXi         {\ensuremath{\Xi}\xspace}                 
 \def\PLambda     {\ensuremath{\Lambda}\xspace}                 
 \def\PSigma      {\ensuremath{\Sigma}\xspace}                 
 \def\POmega      {\ensuremath{\Omega}\xspace}                 
 \def\PUpsilon    {\ensuremath{\Upsilon}\xspace}
 \let\oldPi\Pi
 \def\PPi         {\ensuremath{\oldPi}\xspace}

 \def\PB      {\ensuremath{\mathrm{B}}\xspace}                 
 \def\PD      {\ensuremath{\mathrm{D}}\xspace}                 

 \def\PK      {\ensuremath{\mathrm{K}}\xspace}                 
 \def\Pb      {\ensuremath{\mathrm{b}}\xspace}                 
 \def\Pc      {\ensuremath{\mathrm{c}}\xspace}                 
 \def\Pd      {\ensuremath{\mathrm{d}}\xspace}                 

 \def\Pp      {\ensuremath{\mathrm{p}}\xspace}                 

 \def\Ps      {\ensuremath{\mathrm{s}}\xspace}                 
 \def\Pt      {\ensuremath{\mathrm{t}}\xspace}                 
 \def\Pu      {\ensuremath{\mathrm{u}}\xspace}                 
 \def\thebaroffset{0.0em}
}
{

 \def\Ppi         {\ensuremath{\pi}\xspace}

 \mathchardef\PDelta="7101
 \mathchardef\PXi="7104
 \mathchardef\PLambda="7103
 \mathchardef\PSigma="7106
 \mathchardef\POmega="710A
 \mathchardef\PUpsilon="7107
 \mathchardef\PPi="7105
 \def\PB      {\ensuremath{B}\xspace}                 
 \def\PD      {\ensuremath{D}\xspace}                 

 \def\PK      {\ensuremath{K}\xspace}                 
 \def\Pb      {\ensuremath{b}\xspace}                 
 \def\Pc      {\ensuremath{c}\xspace}                 
 \def\Pd      {\ensuremath{d}\xspace}                 

 \def\Pp      {\ensuremath{p}\xspace}                 

 \def\Ps      {\ensuremath{s}\xspace}                 
 \def\Pt      {\ensuremath{t}\xspace}                 
 \def\Pu      {\ensuremath{u}\xspace}                 
 \def\thebaroffset{0.18em}
}
\newcommand{\offsetoverline}[2][\thebaroffset]{\kern #1\overline{\kern -#1 #2}}%

\makeatletter
\ifcase \@ptsize \relax
  \newcommand{\miniscule}{\@setfontsize\miniscule{4}{5}}
\or
  \newcommand{\miniscule}{\@setfontsize\miniscule{5}{6}}
\or
  \newcommand{\miniscule}{\@setfontsize\miniscule{5}{6}}
\fi
\makeatother

\DeclareRobustCommand{\optbar}[1]{\shortstack{{\miniscule (\rule[.5ex]{1.25em}{.18mm})}
  \\ [-.7ex] $#1$}}

\def\uquark    {{\ensuremath{\Pu}}\xspace}

\def\dquark    {{\ensuremath{\Pd}}\xspace}

\def\squark    {{\ensuremath{\Ps}}\xspace}

\def\cquark    {{\ensuremath{\Pc}}\xspace}

\def\bquark    {{\ensuremath{\Pb}}\xspace}

\def\tquark    {{\ensuremath{\Pt}}\xspace}

\def\pion   {{\ensuremath{\Ppi}}\xspace}
\def\piz    {{\ensuremath{\pion^0}}\xspace}
\def\pip    {{\ensuremath{\pion^+}}\xspace}
\def\pim    {{\ensuremath{\pion^-}}\xspace}

\def\kaon    {{\ensuremath{\PK}}\xspace}
\def\Kbar    {{\ensuremath{\offsetoverline{\PK}}}\xspace}

\def\KorKbar {\kern \thebaroffset\optbar{\kern -\thebaroffset \PK}{}\xspace}
\def\Kz      {{\ensuremath{\kaon^0}}\xspace}
\def\Kzb     {{\ensuremath{\Kbar{}^0}}\xspace}
\def\Kp      {{\ensuremath{\kaon^+}}\xspace}
\def\Km      {{\ensuremath{\kaon^-}}\xspace}

\def\KS      {{\ensuremath{\kaon^0_{\mathrm{S}}}}\xspace}

\def\Kstarm  {{\ensuremath{\kaon^{*-}}}\xspace}

\def\Dbar    {{\ensuremath{\offsetoverline{\PD}}}\xspace}
\def\D       {{\ensuremath{\PD}}\xspace}

\def\DorDbar {\kern \thebaroffset\optbar{\kern -\thebaroffset \PD}\xspace}
\def\Dz      {{\ensuremath{\D^0}}\xspace}
\def\Dzb     {{\ensuremath{\Dbar{}^0}}\xspace}
\def\Dp      {{\ensuremath{\D^+}}\xspace}
\def\Dm      {{\ensuremath{\D^-}}\xspace}

\def\DpDm    {\ensuremath{\Dp {\kern -0.16em \Dm}}\xspace}

\def\Dstarp  {{\ensuremath{\D^{*+}}}\xspace}
\def\Dstarm  {{\ensuremath{\D^{*-}}}\xspace}
\def\Dstarpm {{\ensuremath{\D^{*\pm}}}\xspace}

\def\theDstarp{{\ensuremath{\D^{*}(2010)^{+}}}\xspace}

\def\Dsp     {{\ensuremath{\D^+_\squark}}\xspace}

\def\B       {{\ensuremath{\PB}}\xspace}

\def\BorBbar {\kern \thebaroffset\optbar{\kern -\thebaroffset \PB}\xspace}

\def\Bd      {{\ensuremath{\B^0}}\xspace}

\def\BdorBdbar {\kern \thebaroffset\optbar{\kern -\thebaroffset \Bd}\xspace}

\def\Bs      {{\ensuremath{\B^0_\squark}}\xspace}

\def\BsorBsbar {\kern \thebaroffset\optbar{\kern -\thebaroffset \Bs}\xspace}

\def\Y#1S{\ensuremath{\PUpsilon{(#1S)}}\xspace}

\def\proton      {{\ensuremath{\Pp}}\xspace}

\def\Lz          {{\ensuremath{\PLambda}}\xspace}

\def\LorLbar     {\kern \thebaroffset\optbar{\kern -\thebaroffset \PLambda}\xspace}

\newcommand{\decay}[2]{\mbox{\ensuremath{#1\!\to #2}}\xspace} 

\def\to                 {\ensuremath{\rightarrow}\xspace}

\def\grpsuthree {{\ensuremath{\mathrm{SU}(3)}}\xspace}

\def\order   {{\ensuremath{\mathcal{O}}}\xspace}

\def\eps   {{\ensuremath{\varepsilon}}\xspace}

\def\CP                {{\ensuremath{C\!P}}\xspace}

\newcommand{\Delm}{{\mbox{$\Delta m $}}\xspace}
\newcommand{\ACP}{{\ensuremath{{\mathcal{A}}^{\CP}}}\xspace}

\def\AT#1     {\ensuremath{A_{\mathrm{T}}^{#1}}\xspace}           

\def\C#1      {\ensuremath{\mathcal{C}_{#1}}\xspace}                       
\def\Cp#1     {\ensuremath{\mathcal{C}_{#1}^{'}}\xspace}                    
\def\Ceff#1   {\ensuremath{\mathcal{C}_{#1}^{\mathrm{(eff)}}}\xspace}        
\def\Cpeff#1  {\ensuremath{\mathcal{C}_{#1}^{'\mathrm{(eff)}}}\xspace}       
\def\Ope#1    {\ensuremath{\mathcal{O}_{#1}}\xspace}                       
\def\Opep#1   {\ensuremath{\mathcal{O}_{#1}^{'}}\xspace}                    

\newcommand{\aunit}[1]{\ensuremath{\text{\,#1}}}       

\newcommand{\tev}{\aunit{Te\kern -0.1em V}\xspace}
\newcommand{\gev}{\aunit{Ge\kern -0.1em V}\xspace}
\newcommand{\mev}{\aunit{Me\kern -0.1em V}\xspace}
\newcommand{\kev}{\aunit{ke\kern -0.1em V}\xspace}
\newcommand{\ev}{\aunit{e\kern -0.1em V}\xspace}
 
\newcommand{\mevc}{\ensuremath{\aunit{Me\kern -0.1em V\!/}c}\xspace}
\newcommand{\gevc}{\ensuremath{\aunit{Ge\kern -0.1em V\!/}c}\xspace}
\newcommand{\mevcc}{\ensuremath{\aunit{Me\kern -0.1em V\!/}c^2}\xspace}
\newcommand{\gevcc}{\ensuremath{\aunit{Ge\kern -0.1em V\!/}c^2}\xspace}

\def\fb   {\ensuremath{\aunit{fb}}\xspace}
\def\invfb   {\ensuremath{\fb^{-1}}\xspace}

\def\order{{\ensuremath{\mathcal{O}}}\xspace}
\newcommand{\chisq}{\ensuremath{\chi^2}\xspace}

\def\deriv {\ensuremath{\mathrm{d}}}

\def\gsim{{~\raise.15em\hbox{$>$}\kern-.85em
          \lower.35em\hbox{$\sim$}~}\xspace}
\def\lsim{{~\raise.15em\hbox{$<$}\kern-.85em
          \lower.35em\hbox{$\sim$}~}\xspace}

\def\pt         {\ensuremath{p_{\mathrm{T}}}\xspace}

\newcommand{\lum} {\ensuremath{\mathcal{L}}\xspace}

\def\evtgen     {\mbox{\textsc{EvtGen}}\xspace}

\def\geant      {\mbox{\textsc{Geant4}}\xspace}

\def\photos     {\mbox{\textsc{Photos}}\xspace}

\def\pythia     {\mbox{\textsc{Pythia}}\xspace}

\def\tell1  {TELL1\xspace}
\def\ukl1   {UKL1\xspace}

\newcommand{\lhcborcid}[1]{\href{https://orcid.org/#1}{\hspace*{0.1em}\raisebox{-0.45ex}{\includegraphics[width=1em]{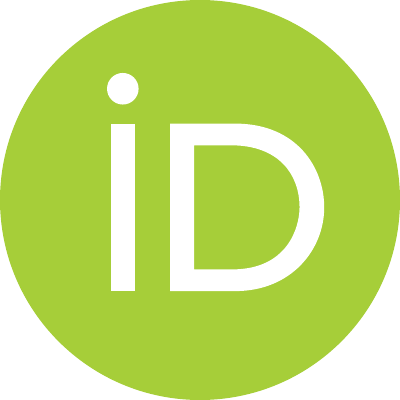}}}}

\hypersetup{
  colorlinks   = true, 
  urlcolor     = blue, 
  linkcolor    = blue, 
  citecolor    = red   
}

\ifEnableSectionTOCLinks
    \usepackage[explicit]{titlesec} 
    
    \let\oldcontentsline\contentsline
    \renewcommand

    \titleformat{\section}{\normalfont\Large\bf}{\hyperlink{tocsection.\thesection}{{\thesection} \parbox[t]{\dimexpr\textwidth-1pc}{#1}}}{1pc}{}

    \titleformat{\subsection}{\normalfont\bf}{\hyperlink{tocsubsection.\thesubsection}{{\thesubsection} \parbox[t]{\dimexpr\textwidth-1pc}{#1}}}{1pc}{}

    \titleformat{name=\section,numberless}[display]{}{}{0pt}{\normalfont\Huge\bfseries #1}
\fi

\usepackage{cite} 
\usepackage{mciteplus}
\usepackage{longtable}

\begin{document}

\renewcommand{\thefootnote}{\fnsymbol{footnote}}
\setcounter{footnote}{1}

\begin{titlepage}
\pagenumbering{roman}

\vspace*{-1.5cm}
\centerline{\large EUROPEAN ORGANIZATION FOR NUCLEAR RESEARCH (CERN)}
\vspace*{1.5cm}
\noindent
\begin{tabular*}{\linewidth}{lc@{\extracolsep{\fill}}r@{\extracolsep{0pt}}}
\ifthenelse{\boolean{pdflatex}}
{\vspace*{-1.5cm}\mbox{\!\!\!\includegraphics[width=.14\textwidth]{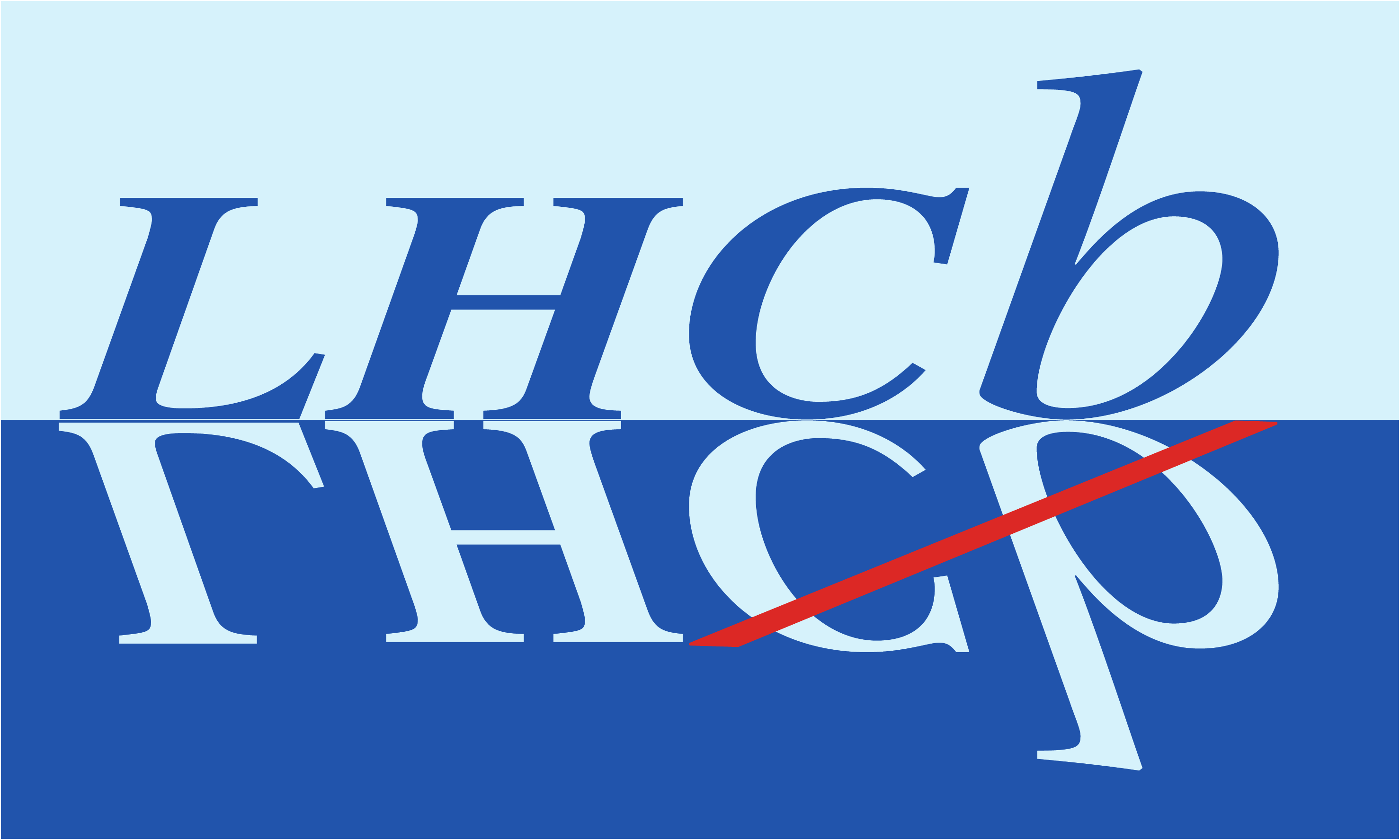}} & &}%
{\vspace*{-1.2cm}\mbox{\!\!\!\includegraphics[width=.12\textwidth]{figs/lhcb-logo.pdf}} & &}%
\\
 & & CERN-EP-2025-221 \\  
 & & LHCb-PAPER-2025-036 \\  
 & & March 3, 2026 \\ 
 & & \\
\end{tabular*}

\vspace*{4.0cm}

{\normalfont\bfseries\boldmath\huge
\begin{center}
  \papertitle 
\end{center}
}

\vspace*{2.0cm}

\begin{center}
\paperauthors\footnote{Authors are listed at the end of this paper.}
\end{center}

\vspace{\fill}

\begin{abstract}
\noindent
A measurement of \CP asymmetry in \decay{\Dz}{\KS\KS} decays is reported, based on a data sample of proton-proton collisions collected with the LHCb detector in 2024 at a centre-of-mass energy of $13.6\tev$, corresponding to an integrated luminosity of $6.2\invfb$.
The $\decay{\Dz}{\KS\pip\pim}$ decay is used as calibration channel to cancel residual detection and production asymmetries. The  time-integrated \CP asymmetry for the \decay{\Dz}{\KS\KS} mode is measured to be
$$\ACP(\decay{\Dz}{\KS\KS}) = (1.86 \pm 1.04\pm 0.41)\% , $$
where the first uncertainty is statistical, and the second is systematic.
This measurement represents the most precise single-experiment determination of this quantity to date.
\end{abstract}

\vspace*{2.0cm}

\begin{center}
  Published in JHEP 02 (2026) 253
\end{center}

\vspace{\fill}

{\footnotesize 
\centerline{\copyright~\papercopyright. \href{\paperlicenceurl}{\paperlicence}.}}
\vspace*{2mm}

\end{titlepage}


\newpage
\setcounter{page}{2}
\mbox{~}
%
%
%
%

\renewcommand{\thefootnote}{\arabic{footnote}}
\setcounter{footnote}{0}

\cleardoublepage

\pagestyle{plain}
\setcounter{page}{1}
\pagenumbering{arabic}

\section{Introduction}
\label{sec:Introduction}
 
Violation of charge-parity (\CP) symmetry in charm-quark transitions has been observed for the first time in the measurement of the difference between the \CP asymmetry in \decay{\Dz}{\Kp \Km} and $\decay{\Dz}{\pip \pim}$ decays~\cite{LHCb-PAPER-2019-006}. As of today, this measurement remains the only experimental evidence of \CP violation in charm decays, and more generally in any up-type quark (\uquark, \cquark, or \tquark) transition.
A successive measurement in the \decay{\Dz}{\Kp \Km} decay mode found no statistically significant \CP asymmetry~\cite{LHCb-PAPER-2022-024}, leaving the $\decay{\Dz}{\pip \pim}$ decay as the only charm-hadron decay with experimental evidence of \CP violation~\cite{LHCb-CONF-2024-004,LHCb-PAPER-2021-033}.

While the existence of \CP violation in the charm-quark sector at the observed $\order(10^{-3})$ level is not excluded in the Standard Model (SM), uncertainties in theoretical calculations do not allow for a quantitative comparison or precise predictions for other decay channels~\cite{Grossman:2012ry,Franco:2012ck,Muller:2015rna,Khodjamirian:2017zdu,Cheng:2019ggx,Grossman:2019xcj,Cheng:2021yrn,Schacht:2021jaz,Gavrilova:2023fzy,Pich:2023kim,Lenz:2023rlq,Bolognani:2024zno}; see Refs.~\cite{Lenz:2020awd,Pajero:2022vev,Petrov:2024ujw} for reviews on this subject. 
Further experimental measurements of \CP asymmetries in charm decays are therefore highly desirable, as they would provide deeper insight into \CP violation phenomenology. Such measurements could potentially involve dynamics beyond the SM~\cite{Grossman:2006jg,Altmannshofer:2012ur,Dery:2019ysp,Bause:2020obd,Buras:2021rdg,Bause:2022jes,Iguro:2024uuw}, that are not constrained to be the same as in the down-type quark (\dquark, \squark, or \bquark) sector. 

Among many charm-hadron decay modes in which \CP violation could manifest itself, the Cabibbo-suppressed \decay{\Dz}{\KS \KS} mode stands out for being analogous to the two-body decays $\decay{\Dz}{\Kp\Km}$ and $\decay{\Dz}{\pip\pim}$~\cite{LHCb-PAPER-2019-006}, but has the possibility of a more sizeable asymmetry. Only amplitudes proceeding via tree-level exchange -- which vanish in the flavour-\grpsuthree limit --  and electroweak loop diagrams contribute to this decay, which therefore has a relatively small branching fraction~\cite{PDG2024}. 
Its \CP asymmetry is  defined by 
\begin{equation}\label{eq:acp_def}
\ACP(\KS\KS) \equiv \dfrac{\Gamma (\decay{\Dz}{\KS\KS}) - \Gamma (\decay{\Dzb}{\KS \KS})}{\Gamma (\decay{\Dz}{\KS\KS}) + \Gamma (\decay{\Dzb}{\KS \KS})},
\end{equation}
with $\Gamma$ being the decay width of the neutral \D meson. Any effect arising from regeneration, mixing and \CP violation in the \KS\KS final state cancel in \ACP, since the \Dz and \Dzb decay to the same \CP-symmetric \mbox{$\Kz - \Kzb$} final state~\cite{Enz:1965tr}.
In the SM, $\ACP(\KS\KS)$ might be significantly larger than in the \decay{\Dz}{\pip\pim} and \decay{\Dz}{\Kp\Km} channels. Several predictions exist for $\ACP(\KS\KS)$~\cite{Brod:2011re,Li:2012cfa,Nierste:2015zra,Cheng:2019ggx,Buccella:2019kpn,Fleischer:2025zhl}, up to the percent level. The study of the \decay{\Dz}{\KS\KS} decay mode provides a valuable probe of \CP violation in the charm sector, offering a means to discriminate between competing theoretical approaches.

These considerations have motivated several experimental determinations of this observable across different experiments, including some recent ones~\cite{CLEO:2000opx,LHCb-PAPER-2015-030,LHCb-PAPER-2020-047,CMS:2024hsv,Belle:2024vho,Belle-II:2025ezw}.
The world average of the \CP asymmetry is $\ACP (\KS\KS) = (-1.19 \pm 0.77 \pm 0.17)\%$, where the time-dependent contribution to the asymmetry is neglected~\cite{charm-fitter, Pajero:2021jev}. This has now reached the subpercent level, entering the interesting sensitivity range of theoretical predictions.
The reduced precision of this measurement compared with other charm \CP asymmetry results is due to the smaller branching fraction and the difficulty of reconstructing the \KS decay products, given its much longer lifetime relative to bottom and charm hadrons.

In this work, a new measurement of $\ACP (\KS\KS)$ is reported, based on a sample of proton-proton ($pp$) collisions corresponding to an integrated luminosity of about $6.2\,\invfb$, collected in 2024 by the LHCb detector. 
The detector has been significantly improved over its previous version, including the data acquisition system~\cite{LHCb-TDR-016, LHCb-TDR-021}, which enabled the adoption of a software-based selection of \KS particles at the first trigger level, resulting in a substantially improved efficiency. 

The measurement of $\ACP(\KS\KS)$  requires knowledge of the \Dz flavour at production.
A sample of flavour-tagged $\decay{\Dz}{\KS \KS}$ decays is obtained by selecting only \Dz mesons that originate from $\decay{\theDstarp}{\Dz \pip}$ decays.\footnote{Inclusion of the charge-conjugate process is implied throughout this document unless explicitly specified. Hereafter, the \theDstarp meson is referred to as \Dstarp.} 
The charge of the pion in this decay (``tagging pion'') identifies the flavour of the accompanying neutral charm meson. While $\Dz-\Dzb$ oscillations can cause a flavour change before decay, the effect is small compared with the resolution of the current measurement and is therefore neglected~\cite{LHCb-PAPER-2024-008}. The \KS mesons are reconstructed in the $\pip\pim$ final state and are required to have decayed within the vertex detector.

The decay widths $\Gamma$ in Eq.~\ref{eq:acp_def} are related to the observed yields $N$ by 
\begin{equation}
N(\decay{\DorDbar^0}{\KS \KS}) \propto \eps^{\pm}(\Dstarpm) \sigma (\Dstarpm)\,\Gamma (\decay{\DorDbar^0}{\KS \KS}),
\end{equation}
where $\sigma(\Dstarpm)$ denotes the production cross-sections of the \Dstarpm mesons and $\eps^{\pm}(\Dstarpm)$ the corresponding detection efficiencies for their decays.
Both factors are charge asymmetric, due to the \Dstarpm production asymmetry arising from the hadronisation of charm quarks in $pp$ collisions, and to asymmetries in the geometry and response of the detector.
Since the final state of the \Dz decay is self-conjugate, any detection asymmetry arises solely from the tagging pion. 
All these quantities are calibrated using a sample of $\decay{\Dz}{\KS \pi^+\pi^-}$ decays, which have the same final-state particles as the $\decay{\Dz}{\KS\KS}$ decay, but with a negligible \CP asymmetry~\cite{Pajero:2022vev}.
In order to avoid experimenter’s bias, the results of the analysis were not examined until the full procedure had been finalised.

\section{LHCb detector}

The LHCb Run~3 detector~\cite{LHCb-DP-2014-002} is a single-arm forward spectrometer covering the pseudorapidity range $2 < \eta < 5$, designed for the study of particles containing \bquark\ or \cquark\ quarks. Installed prior to the start of Run~3 data taking in 2022, it represents a major upgrade of the detector system. The new configuration was designed to match the performance of the Run 1--2 detector~\cite{LHCb-DP-2022-002,LHCb-DP-2008-001}, while allowing it to operate at approximately five times the luminosity.  

The high-precision tracking system consists 
of a silicon-pixel vertex detector (VELO) surrounding the $pp$
interaction region~\cite{LHCb-TDR-013}, a large-area silicon-strip detector (UT)~\cite{LHCb-TDR-015} located upstream of a dipole magnet with a bending power of about $4{\mathrm{\,T\,m}}$, and three stations of scintillating-fibre detectors~\cite{LHCb-TDR-015} located downstream of the magnet.
Tracks reconstructed in both the VELO and the scintillating-fiber detectors, with the possible addition of intermediate hits in UT, are referred to as {\em long} tracks.
The dipole magnetic-field polarity is periodically reversed during data taking, alternating between pointing upwards (\MagUp) and downwards (\MagDown), to mitigate the effects of differences in reconstruction efficiencies for particles of opposite charges.
Different types of charged hadrons are distinguished using information from two ring-imaging Cherenkov detectors~\cite{LHCb-TDR-014}.

Readout of all detectors into an all-software trigger~\cite{LHCb-TDR-016,LHCb-TDR-021} is a central feature of the upgraded detector, allowing the reconstruction of events at the maximum LHC interaction rate, and their selection in real time. The trigger system is implemented in two stages: a first inclusive stage (HLT1) running on GPU boards based
primarily on charged particle reconstruction, which reduces the data volume by
roughly a factor of twenty, and a second stage (HLT2) running on CPUs, which performs the full offline-quality reconstruction and selection of physics signatures. A large disk buffer is placed between these stages to hold the data while the real-time alignment and calibration are performed.

For the measurement described in this paper, data collection benefits from the newly implemented trigger system, which, for the first time, enables the selection of events containing \KS candidates directly at the HLT1 level for every LHC bunch crossing. This leads to an improvement in the signal efficiency by a factor of about three over the previous \lhcb detector, which is the main factor responsible for the increase in precision for the current measurement. 

Simulated samples are used solely to optimise selection requirements. In the simulation, $pp$ collisions are generated using
  \pythia~\cite{Sjostrand:2007gs} with a specific
  \lhcb configuration~\cite{LHCb-PROC-2010-056}.
Decays of unstable particles are described by \evtgen~\cite{Lange:2001uf}, in which final-state radiation is generated using \photos~\cite{davidson2015photos}.
The interaction of the generated particles with the detector, and its response, are implemented using the \geant toolkit~\cite{Allison:2006ve} as described in Ref.~\cite{LHCb-PROC-2011-006}.

\section{Event selection}

At the \hltone stage, long tracks with good fit $\chisq$, momentum $p>5\gevc$, transverse momentum $\pt>0.45\gevc$, and incompatible with originating from any reconstructed primary vertex (PV) are considered. The tracks with opposite charge passing this selection are paired, and their invariant mass is computed assuming the pion mass for each of them. Pairs with an invariant mass within $\pm 45\mevcc$ of the known \KS mass~\cite{PDG2024} are classified as \KS candidates. Events of interest for this analysis are triggered by requiring at least one \KS candidate with $\pt >2.45\gevc$.

At the \hlttwo level, after a full reconstruction of the whole event, \KS candidates with an invariant mass within $\pm 35\mevcc$ from the known \KS mass are paired to form \Dz candidates. These candidates are required to originate from a common vertex and to have an invariant mass within $\pm 90\mevcc$ of the known \Dz mass~\cite{PDG2024}. An additional track is then combined with the \Dz particle to form \Dstarp candidates. The invariant-mass difference $\Delm= m(\Dstarp)-m(\Dz)$ is required to be smaller than $170\mevcc$~\cite{PDG2024}. The resulting candidates are saved on permanent storage and used for further analysis.

A calibration sample of $\decay{\Dstarp}{\Dz(\KS\pip\pim)\pip}$ decays is obtained from the same \hltone trigger selection, by applying a similar \hlttwo selection, in which the requirements on the second \KS of the signal mode are replaced by requirements on the compatibility of the two pions to have originated from the \Dz decay vertex. This \hlttwo selection is specifically developed for the purpose of the current work, with the aim of providing a calibration sample with the closest trigger selections to that of the signal decay. Events collected via other parallel trigger selections are not considered. This ensures an accurate cancellation of detector effects between samples across all periods of data taking, notwithstanding the changes in detector and trigger conditions made over time 
to optimize the global apparatus performance. 

The offline selection of candidate decays is designed to suppress the background while avoiding introducing any charge biases that might affect the signal and calibration mode differently. 
In this respect, an important aspect is the presence, in both the signal and calibration sample, of a non-negligible fraction of charm mesons originating from \bquark-hadron decays rather than from the primary \proton \proton interaction. These secondary particles are affected by different production asymmetries with respect to the prompt decays, quantified at the 0.2\% level~\cite{LHCb-PAPER-2020-045}. 
Following the same approach as the previous LHCb measurement~\cite{LHCb-PAPER-2020-047}, it has not been attempted to reject secondary decays,
but rather allow them to contribute to the measurement.
The spurious asymmetry introduced by the secondary component is cancelled by ensuring that the calibration sample contains secondary decays in the same proportion as the signal sample. 
This is achieved by a careful choice of the selection criteria, avoiding imposing any requirements on observables sensitive to the position of the \Dstarp or \Dz vertices that have different resolutions in the two samples~\cite{phd-thesis-pica}.

The calibration mode has the same final-state particles as the signal, and a branching fraction approximately two hundred times larger~\cite{PDG2024}. This decay mode can contaminate the signal even though the proportion of pion pairs randomly falling within the \KS mass window is small. Although the fit procedure is capable of taking this contribution into account, additional selection criteria are applied on the significance of the flight distance of both \KS mesons from the \Dz decay vertex to suppress this contamination to a negligible level. The effect of this requirement is independent of the flight distance of the parent charm meson from the primary \proton \proton interaction vertex, and therefore does not affect the proportion of the secondary component.
Studies based on simulated events have identified other physics processes that could potentially contaminate the signal sample, such as partially reconstructed \Dz decays originating from \Dstarp meson, including $\decay{\Dz}{\KS \KS} \piz$ decays. Some non-\Dz decays can also contaminate the sample, with the main contributor being the $\decay{\Dsp}{\KS \KS \pip}$ mode, where the $\KS \KS$ pair is incorrectly associated with a $\Dz$ decay. These background components are rendered negligible by selecting only candidates with $m(\Dz)$ within $\pm 26\mevcc$ of the known \Dz mass~\cite{PDG2024}, which is measured with a resolution of about $7\mevcc$. The width of the mass window is chosen to prevent a possible risk of charge biases arising from residual differences in the mass calibration and resolution between positive and negative particles. The potential contamination of the \KS sample by \Lz particles has been verified on data, and found to have a negligible impact, since this background does not peak in the \Delm distribution.

To maximise the sensitivity of the measurement, only \Dstarp candidates compatible with having decayed in the vicinity of the PV are selected. This suppresses background from random combination of tracks and allows the candidate mass to be recalculated by constraining the origin vertex to coincide with the PV, improving the \Delm mass resolution by a factor of two, with an associated improvement in combinatorial background rejection. 
Using simulated samples it is shown that this requirement removes the same proportion of secondary \B-decay component from both the signal and calibration sample, thus leaving the validity of the cancellation procedure unaffected.

The largest remaining background contribution to the signal sample originates from random combinations of genuine \KS particles from the primary interaction, which form a fake \Dz candidate. This is partially rejected by the $m(\Dz)$ requirement, 
and is further suppressed by requiring the \KS candidates to be inconsistent with originating from the PV with a selection on its  impact parameter significance, thereby improving the signal purity.
To ensure an identical effect on the secondary fraction in signal and calibration samples, the selection is applied only on the \KS candidate that fulfils the HLT1 selection. If both \KS candidates do, one is arbitrarily chosen. A threshold for this requirement has been optimised jointly with a condition on a multivariate classifier. Track-related observables, including track and vertex quality, transverse momenta of \KS and \Dz candidates, helicity angles of the \KS and \Dz decays, and particle-identification information of the \Dz final-state particles, are used as input to a $k$-nearest-neighbours (kNN) classifier~\cite{kNN}. This classifier is trained using a simulated $\decay{\Dz}{\KS\KS}$ sample as signal proxy and data from the \Dz mass sidebands as combinatorial background proxy. 
Kinematic variables of the tagging pion are excluded from the classifier to avoid introducing possible charge-asymmetry biases.

To make better use of the available data, signal candidates are classified into two classes of purity (high-purity, with signal-to-background ratio $S/B\sim 20$ and low-purity, with $S/B\sim4$), which are fitted separately to determine \ACP. These classes are defined by an optimised set of requirements on both the kNN classifier and the significance of the \KS impact parameter, chosen to maximise the combined resolution on the \ACP parameter. 

This strategy closely follows the procedure of the previous LHCb publication~\cite{LHCb-PAPER-2020-047}. In about $10\%$ of events, multiple \Dstarp candidates are found. In these cases, only one \Dstarp candidate is arbitrarily selected for subsequent analysis.

The data sample is divided into eight blocks according to data-taking periods with substantially different trigger and detector spatial alignment conditions. These conditions were varied several times during the year to optimise the data acquisition. 
The offline selection is the same in each block, except for the requirement on the \Dz mass window, whose threshold has been determined separately for each block to account for possible slight calibration differences. 
The integrated luminosities of each data block are reported in Table~\ref{tab:yields}. 
    
\begin{table*}[tbp]
	\begin{center}
		\caption{Integrated luminosity of individual data blocks.}
		\label{tab:yields}
	\begin{tabular}{l|c|c}
       Data block & $\int{\lum \deriv t}$ [\invfb] & Polarity \\
       \hline
       A & 0.7 & \MagDown \\
       B & 0.7 & \MagUp \\
       C & 0.6 & \MagUp \\
       D & 1.1 & \MagUp \\
       E & 1.1 & \MagUp \\
       F & 0.9 & \MagDown \\
       G & 0.7 & \MagDown \\
       H & 0.4 & \MagUp \\
       \hline
       Total & 6.2 & --\\
\end{tabular}
\end{center}
\end{table*}

\section{Correction of nuisance asymmetries}

Each \decay{\Dz}{\KS\KS} candidate is appropriately weighted to cancel all spurious production and detection asymmetries, with weights obtained from the calibration sample. This follows the same procedure as the previous \lhcb analysis~\cite{LHCb-PAPER-2020-047}, with the only modification that the weights, $w^\pm$, are expressed in terms of the three-momentum  $\vec{p}$ of the tagging pion from the \Dstarp decay instead of using the \Dz kinematic observables
\begin{equation}\label{eq:weight_def}
    w^\pm(\vec{p}) =  \frac{n_C^+(\vec{p})+n_C^-(\vec{p})}{2 n_C^\pm(\vec{p})}.
\end{equation}
This is motivated by the need to precisely control the most important source of detection asymmetry. In Eq.~\ref{eq:weight_def}, the probability density of calibration decays in the $\vec{p}$ space is denoted by $n_C^\pm(\vec{p})$. The plus and minus signs in the expression indicate quantities related to \Dstarp and \Dstarm candidates, respectively.
The weights are computed by classifying the calibration sample based on the observed charge asymmetry by means of a kNN classifier based on the three-dimensional $\vec{p}$ space. The number of neighbours used is 30, as it is low enough to describe local variations of asymmetry across the phase space, and it allows to keep statistical fluctuations related to the size of the considered sample below $10\%$.

In the calibration sample, the absence of selection criteria requiring compatibility of the \Dz final-state particles with the primary vertex in the dedicated \hlttwo selection leads to a low-purity sample. To improve the $S/B$ ratio, additional offline selections are applied, while maintaining strict consistency between the calibration and signal selections to ensure an accurate cancellation of detector effects.

\begin{figure}[tb]
	\centering
	\includegraphics[width=1.\textwidth]{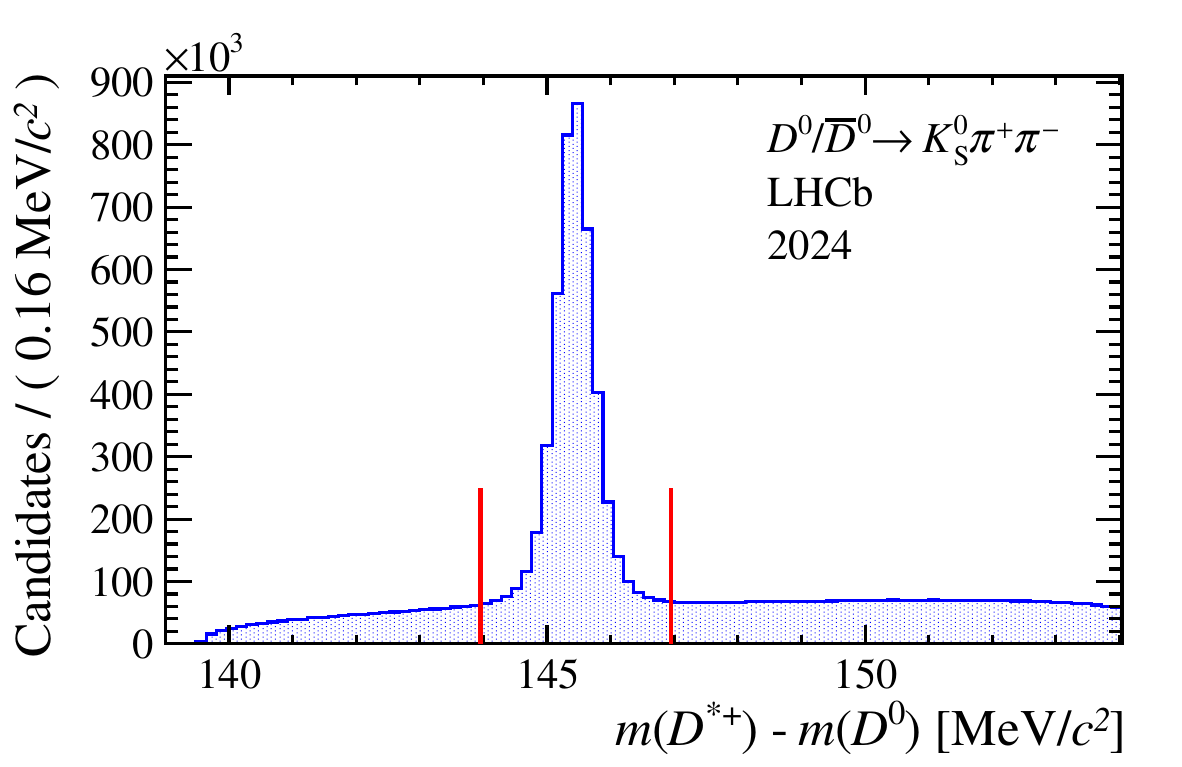}
	\caption{Distribution of $\Delm = m(\Dstarp) - m(\Dz)$ for \decay{\Dz}{\KS\pip\pim} candidates, after application of all offline selections. Calibration candidates are selected by the requirement \mbox{$|\Delm - 145.45\mevcc| < 1.5 \mevcc$}, shown in red.}
    \label{fig:nobias_cuts}
\end{figure}

First, all offline selections applied to the signal sample that can be directly ported to the calibration sample are applied to equalise the kinematic distributions and the fraction of secondary decays. 
Then, requirements on variables including the particle-identification information of final-state tracks, vertex-fit quality of \Dz, \KS and \Dstarp mesons, and the opening angles between the final-state particles are applied to further suppress the combinatorial background.
The resulting sample has a high purity ($S/B\sim 13$) and it is used for the weight calculation. The \Delm distribution of calibration candidates is shown in Fig.~\ref{fig:nobias_cuts}. It is verified on block E data that the potential bias in the measurement due to the residual background under the \Delm peak is less than $10^{-4}$.

The newly adopted calibration mode is not self-conjugate, in contrast to the previously used \decay{\Dz}{\Kp\Km} decay~\cite{LHCb-PAPER-2020-047}. To account for this, the analysis procedure is modified with respect to the earlier \lhcb measurement. The decay proceeds primarily through the intermediate \decay{\Dz}{\Kstarm(\KS \pim)\pip} channel, which produces a charge-asymmetric momentum distribution of the two pions. As a result, the detector response may differ between signal and calibration candidates, since positive and negative pions traverse regions with distinct charge-dependent detection asymmetries in the two samples. To eliminate this spurious effect, the three-momentum of the pions from the \Dz decay is weighted in the calibration sample to make it charge symmetric, before being used as input to the kNN classifier that calculates the signal weights. The charge-symmetrisation procedure employs an independent kNN classifier, trained to distinguish candidates based on the charge of the highest-momentum pion and the kinematics of both final-state pions. The effectiveness of this correction is confirmed by verifying that a kNN classifier can no longer separate \Dz from \Dzb after the weights are applied. The correction shifts the final result by about $0.1\%$, corresponding to roughly $14\%$ of the statistical uncertainty, and the associated systematic uncertainty is therefore negligible compared to other contributions.

\begin{figure}[tb]
    \centering
    \includegraphics[width=1.\textwidth]{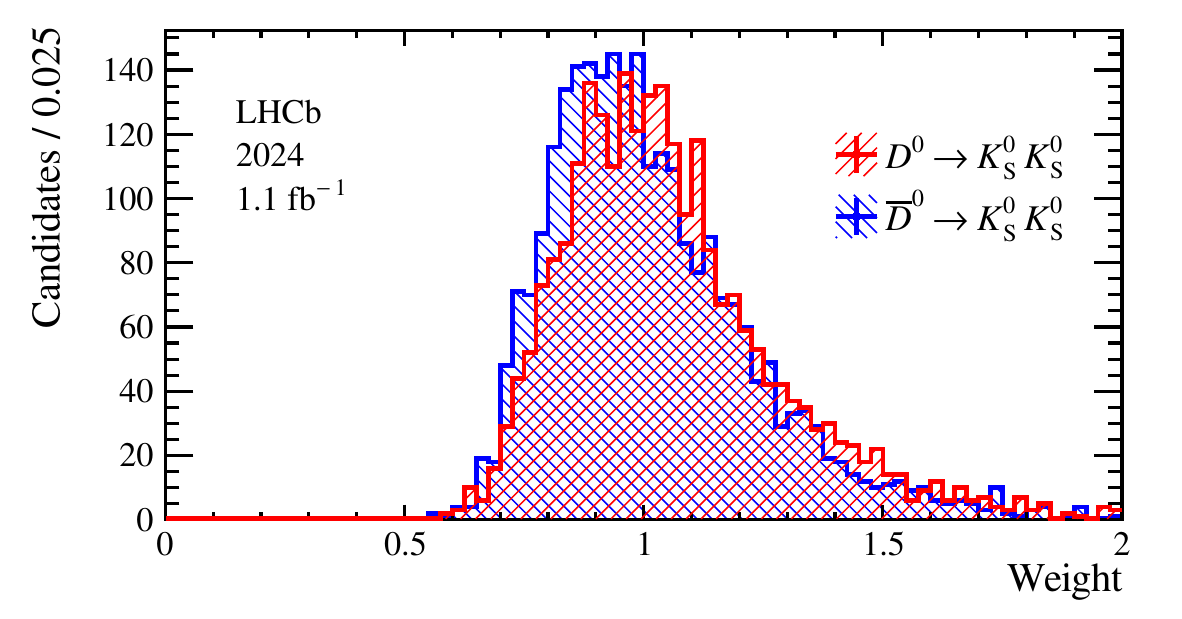}
    \caption{Example (data block D) of distributions of weights applied to signal candidates. These are used in the fit extracting $\ACP(\KS\KS)$ to correct the signal sample for detection and production asymmetries of the calibration channel.}
    \label{fig:weights}
\end{figure}

Distributions of the final weights $w^\pm$ applied to the signal sample for the correction of spurious asymmetries (as defined in Eq.~\ref{eq:weight_def}) are shown in Fig.~\ref{fig:weights}. The differences between the distributions of the two charges are a consequence of different \Dz and \Dzb acceptances and detector asymmetries. These distributions are similar, but not identical, in every data block, and \Dz/\Dzb candidates are separately corrected before fitting for the asymmetry.
To avoid weights affected by large uncertainties, candidates with weights lower than 0.56 and larger than 5.00 are dropped at this stage. These correspond to a fraction of $5\times10^{-4}$, whose effects are therefore neglected. The size of the asymmetry correction obtained by the application of these weights is about 1.4\%.  This accounts for the cumulative effect of all production and detection asymmetries affecting the data.

\section{Asymmetry measurement}
A binned maximum-likelihood fit to the joint distribution of \Delm and the two $m(\KS)$ observables is performed separately for each data block to measure the \ACP parameter. The choice of the \KS candidate for filling the invariant-mass histograms is arbitrary. The three-dimensional fit is performed simultaneously for candidates of both flavours and across the two purity bins. Data block A is further divided into two sections, fitted separately to account for changes in running conditions. 
The final measurement is then obtained as the weighted average of the results from all blocks. 

The total probability density function is parametrised by the sum of the signal component, which has a peaking distribution in each of the three observables, and four background components,  each describing a specific source: random combination of five pions, a \KS combined with three random pions, three random pions combined with a \KS, two \KS combined with a random pion. The probability density functions used to describe the five components are empirical.
The contributions from \decay{\Dz}{\KS\pip\pim} and \decay{\Dz}{\pip\pim\pip\pim} decays are verified to be negligible, and are excluded from the fit.
The peaking component in the \Delm distribution is described by a Johnson $S_U$ distribution~\cite{Johnson:1949zj}. 
The peaking component in the $m(\KS)$ distribution is modelled as the sum of two Gaussian functions with a common mean but different widths, except for blocks A, B, and H, where a single Gaussian function is used to avoid fit instabilities due to their smaller sample sizes. 
The nonpeaking component in the 
\Delm distribution is modelled with a threshold function in the low-purity bins and with a first-order polynomial in the high-purity bins. In the
$m(\KS)$ distribution, this component is described by a first-order polynomial.

In each subsample, the parameters defining the signal and background probability density functions are shared between \Dstarp/\Dstarm candidates, while the normalisation of each component is allowed to differ. All parameters included in the probability density function are allowed to vary in the fit to the data.

\begin{figure}[!h]
	\centering
	\includegraphics[width=0.49\textwidth]{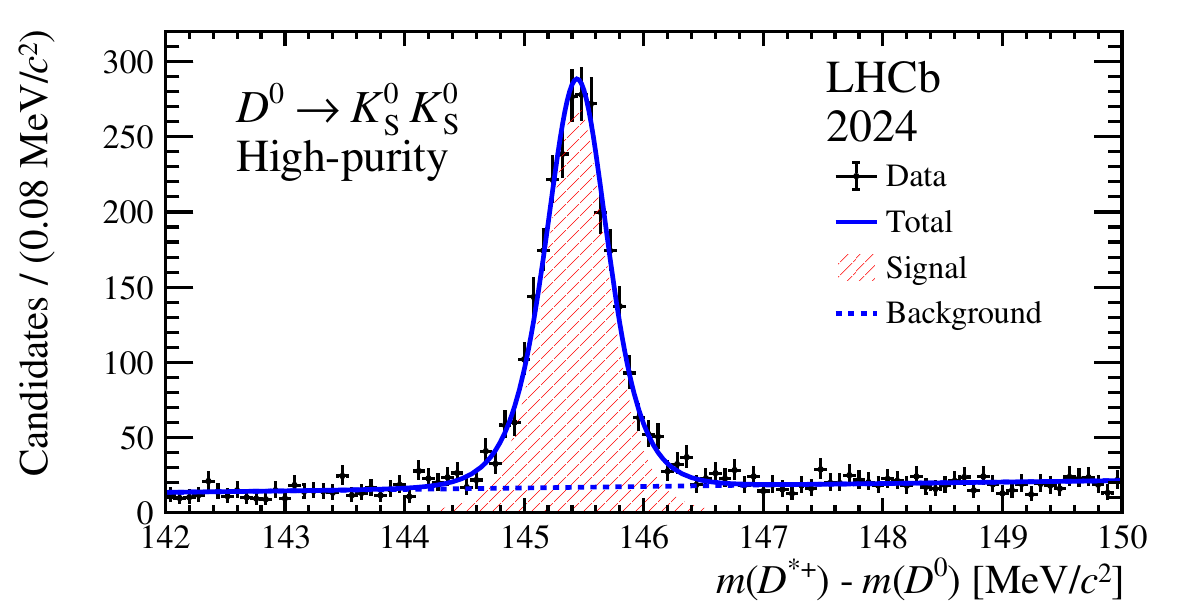}
    \includegraphics[width=0.49\textwidth]{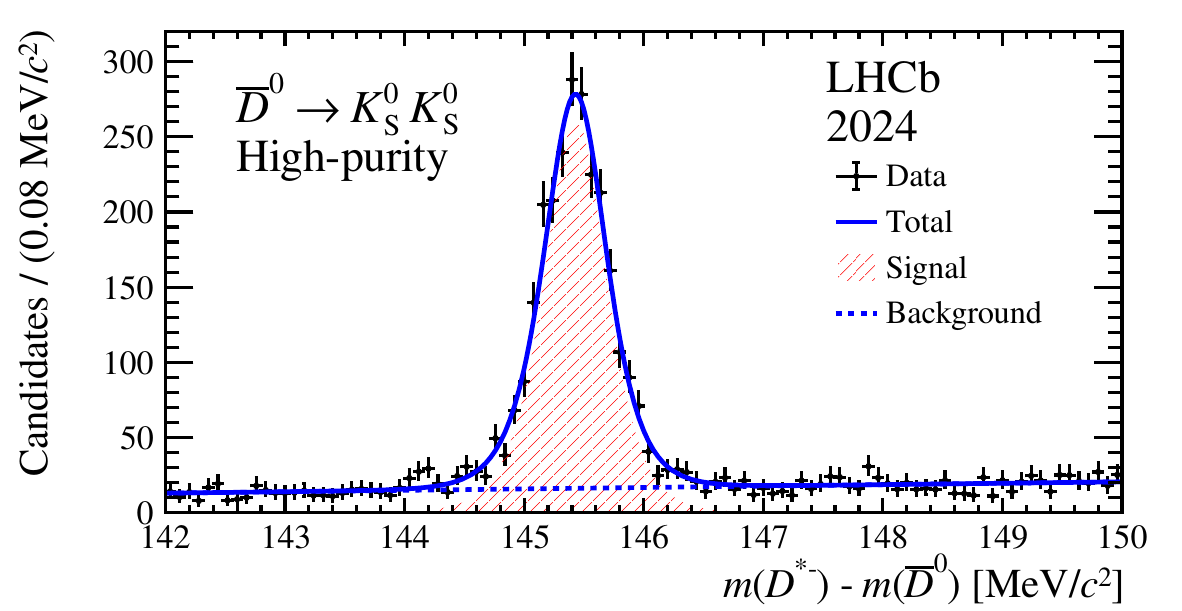}\\
    \includegraphics[width=0.49\textwidth]{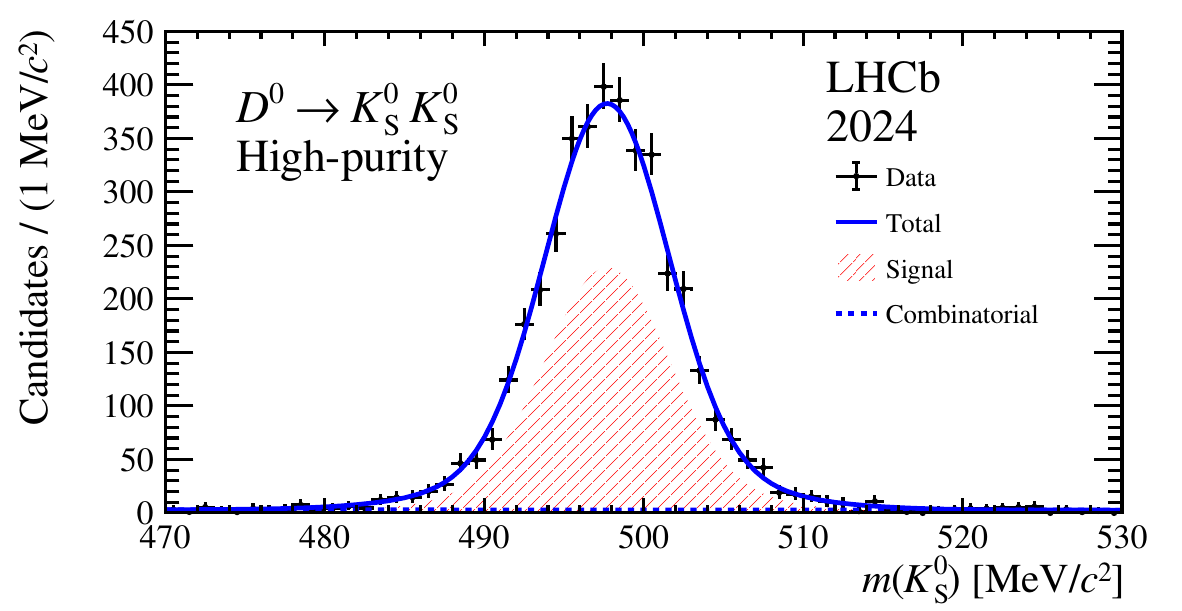}
    \includegraphics[width=0.49\textwidth]{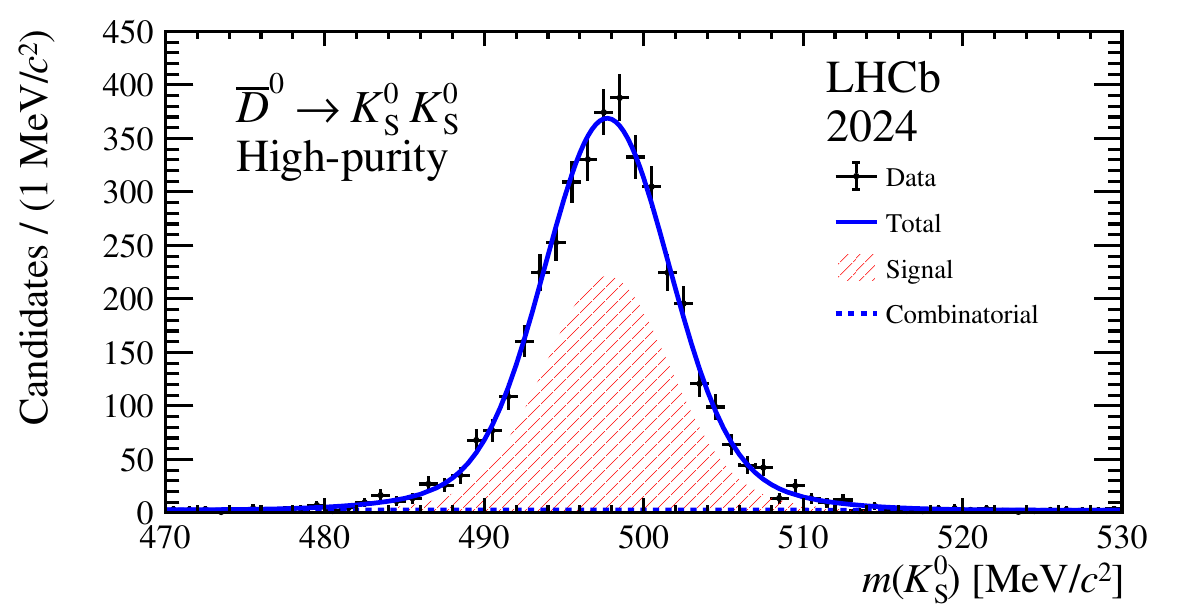}\\
    \includegraphics[width=0.49\textwidth]{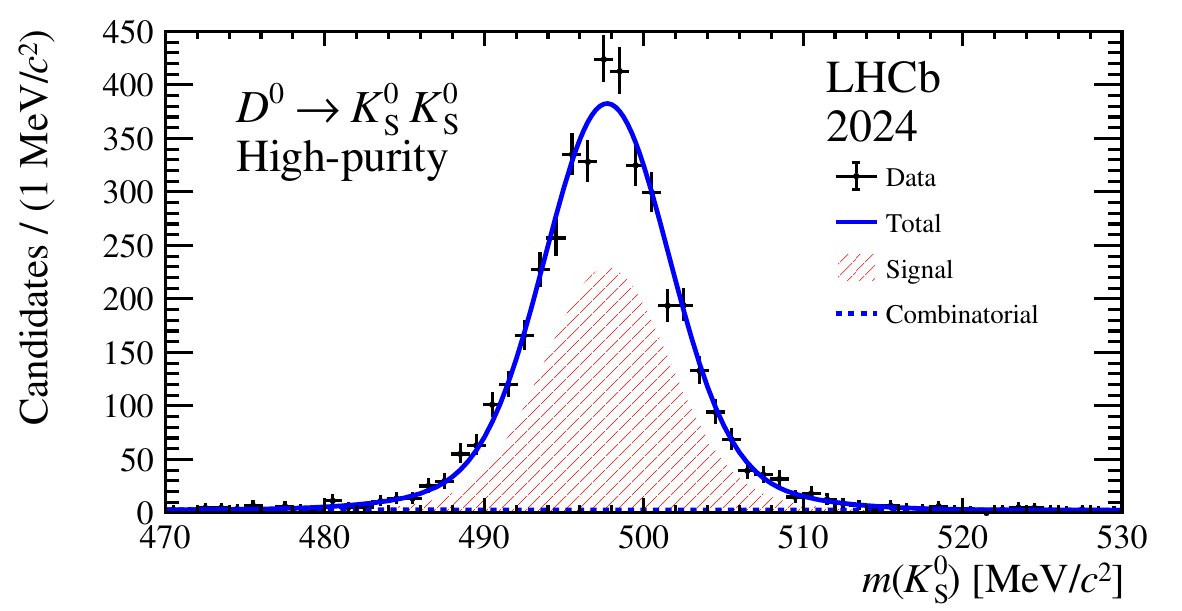}
    \includegraphics[width=0.49\textwidth]{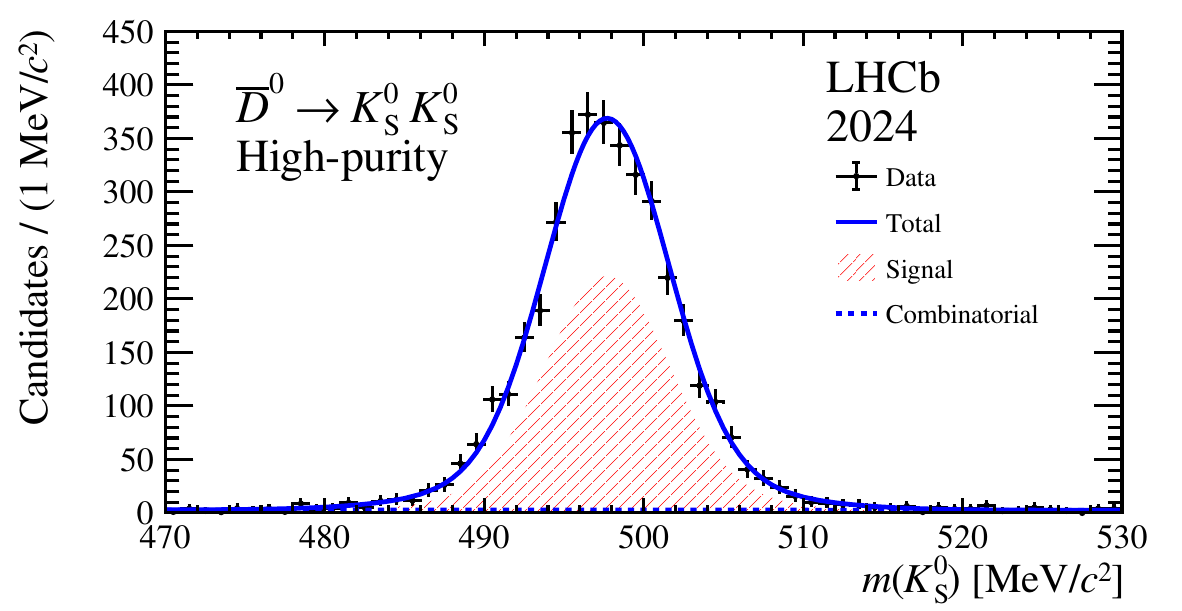}\\
	\caption{Results of the fit for the sample satisfying the high-purity selection and combining all data blocks. The left (right) column reports histograms for the \Dz (\Dzb) candidates. The \Delm distributions are reported in the top row. The masses of the two \KS candidates are reported in the middle and bottom rows. The \decay{\Dz}{\KS\KS} component is indicated by the red area. For the $m(\KS)$ distributions the combinatorial component refers to random combination of two pions for the considered candidate, paired either with a genuine \KS candidate or with an independent random combination of two additional pions.} \label{fig:plots_bin1}
\end{figure}

\begin{figure}[!htb]
	\centering
	\includegraphics[width=0.49\textwidth]{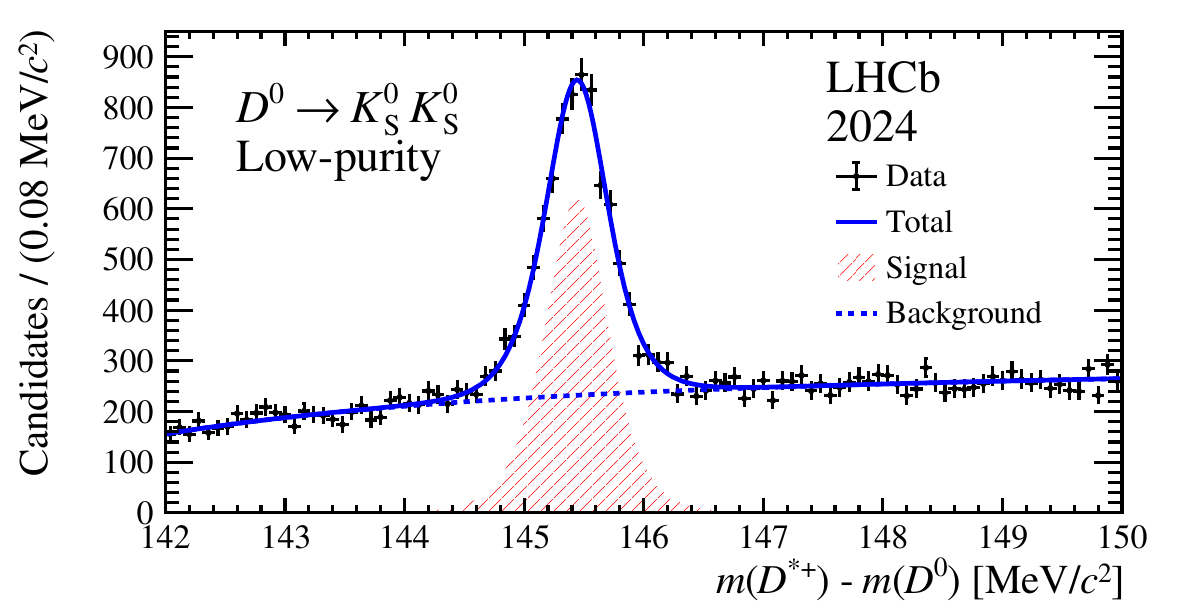}
    \includegraphics[width=0.49\textwidth]{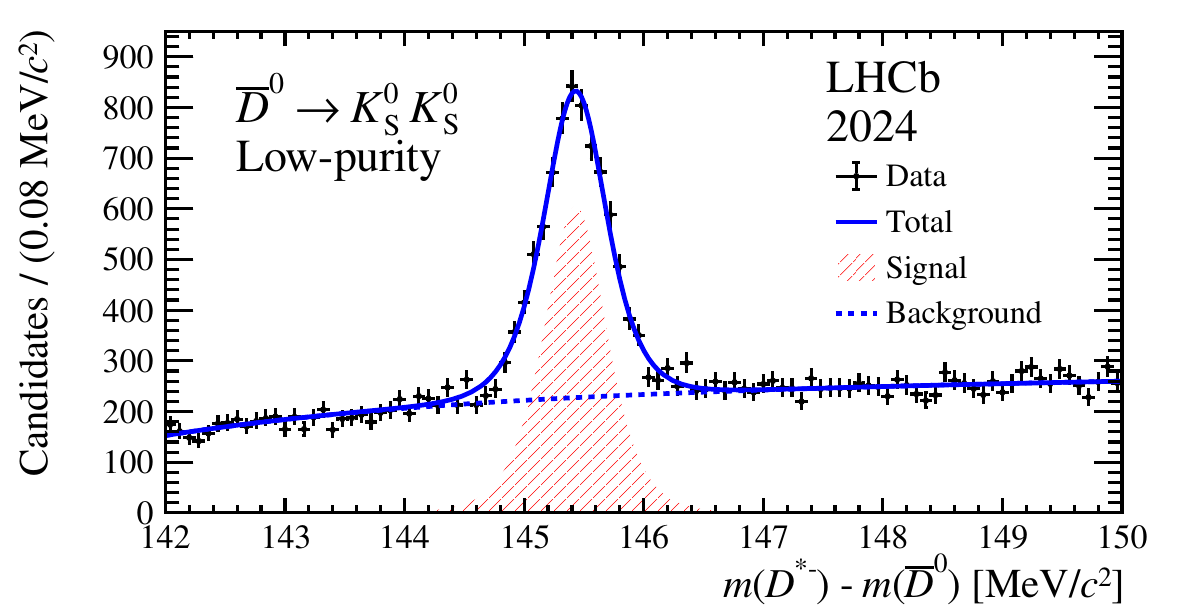}\\
    \includegraphics[width=0.49\textwidth]{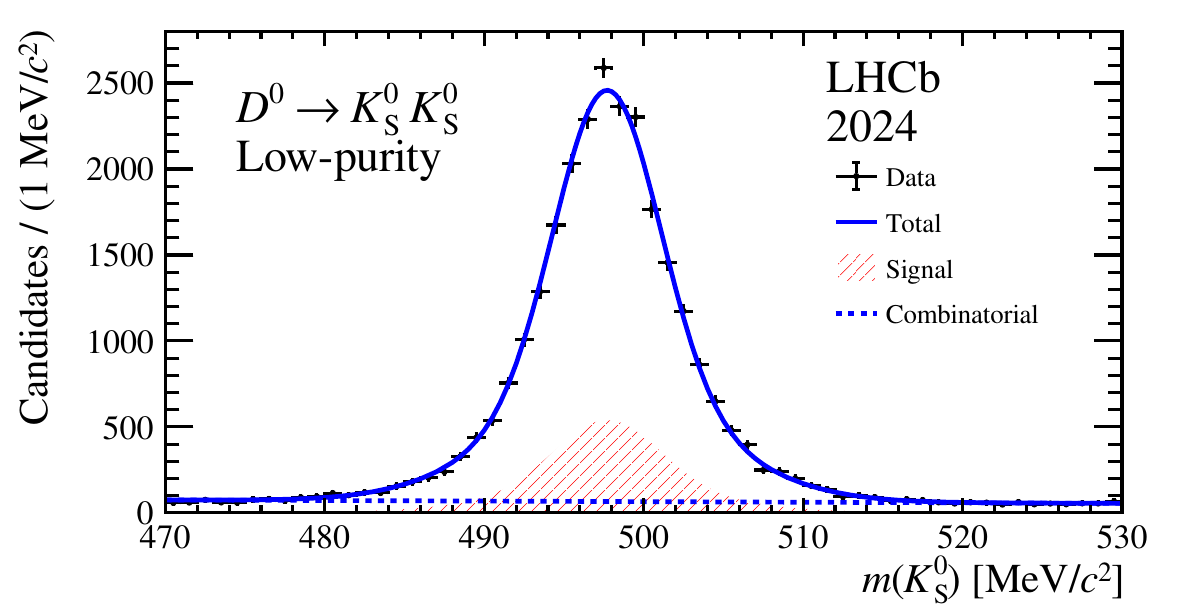}
    \includegraphics[width=0.49\textwidth]{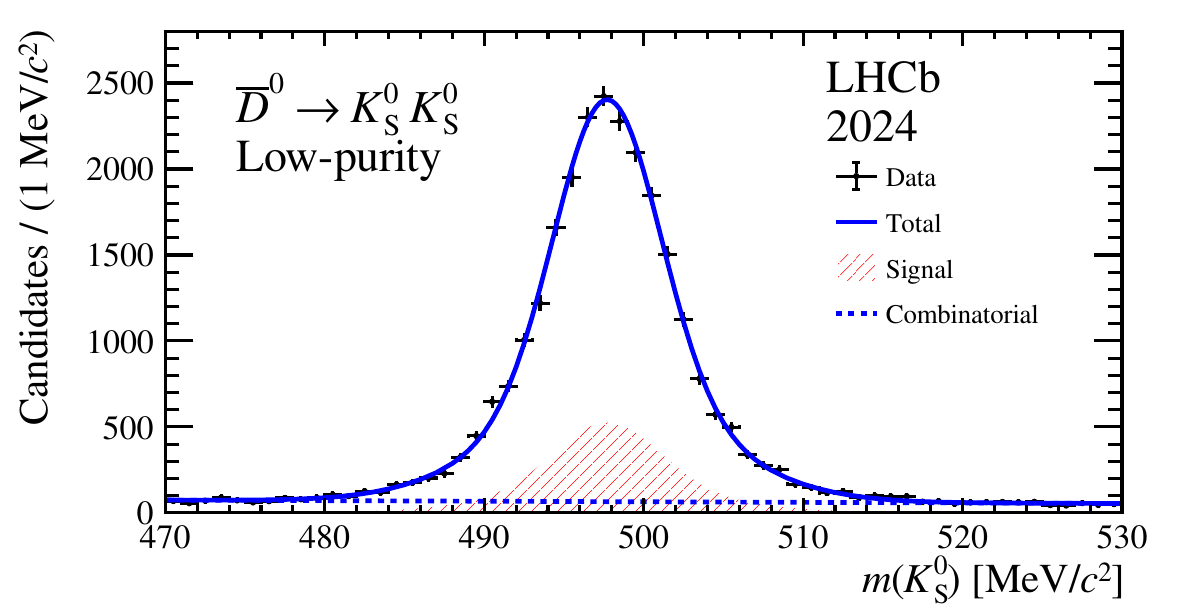}\\
    \includegraphics[width=0.49\textwidth]{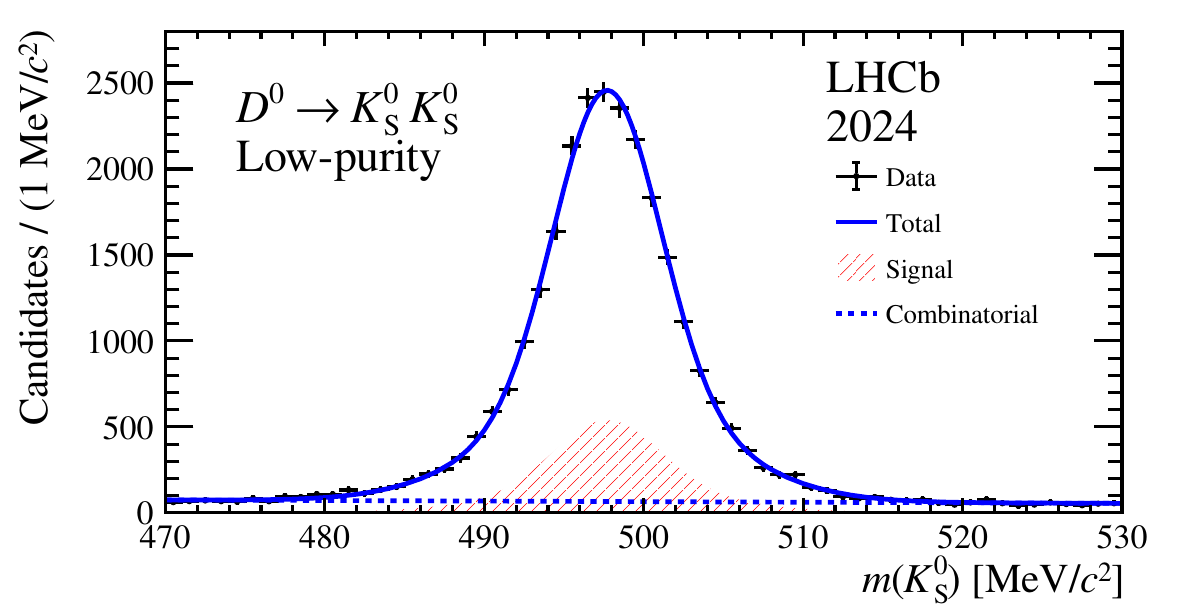}
    \includegraphics[width=0.49\textwidth]{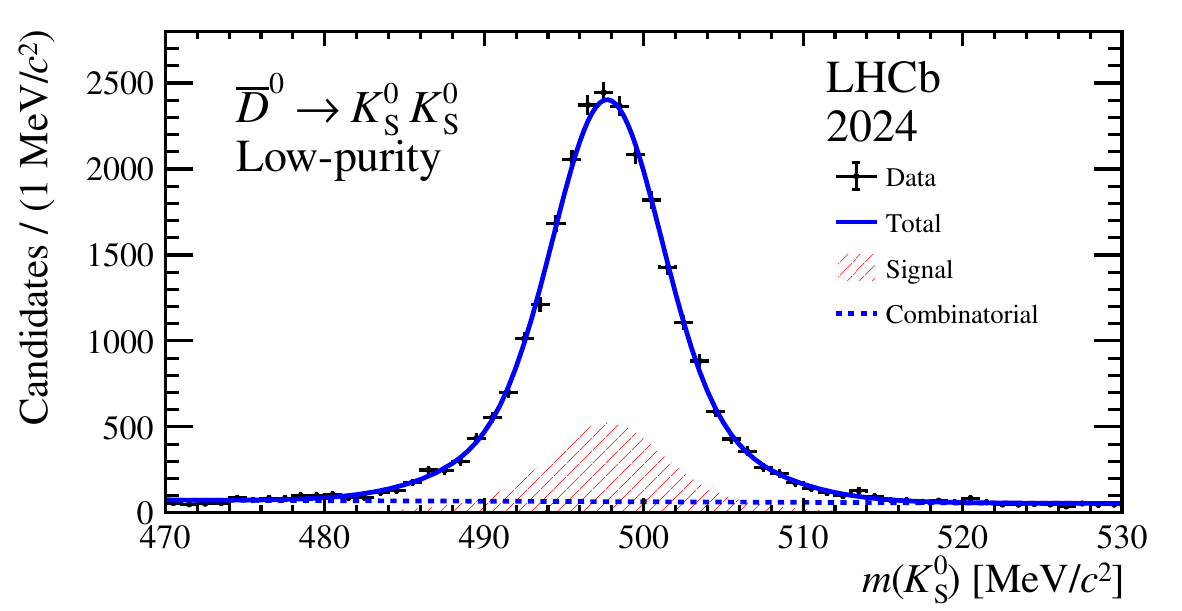}\\
	\caption{Results of the fit for the sample satisfying the low purity selection and integrating all data blocks. The left (right) column reports histograms for the \Dz (\Dzb) candidates. The \Delm distributions are reported in the top row. The masses of the two \KS candidates are reported in the middle and bottom rows. The \decay{\Dz}{\KS\KS} component is indicated by the red area. For the $m(\KS)$ distributions the combinatorial component refers to random combination of two pions for the considered candidate, paired either with a genuine \KS candidate or with an independent random combination of two additional pions.} \label{fig:plots_bin2}
\end{figure}

Figures~\ref{fig:plots_bin1} and \ref{fig:plots_bin2} show the projections of the fit to the total data sample for candidates satisfying the high- and low-purity selection, respectively. The histograms are obtained by adding all data blocks. The candidates entering these plots have been individually weighted to correct for spurious asymmetries.

\section{Systematic uncertainties and cross-checks}

To assess potential systematic effects, the weighting procedure is validated by applying the weights derived from one half of the calibration sample to the other half. As expected, this cancels the raw asymmetry of the sample, which was initially significant at 
$(-1.04\pm0.05)$\% due to a combination of production asymmetry and detector effects, and reduces it to 
$(-0.08\pm0.07)$\% after the application of the weights.

The weighting procedure is also affected by a systematic uncertainty, partly due to the finite size of the calibration sample, and partly due to the numerical approximation, which is implicit in estimating a ratio of probability densities with the discretised kNN classifier output. 
The first uncertainty is evaluated by a bootstrap procedure applied to the calibration sample, where multiple pseudoexperiments are generated by resampling. The resulting uncertainty is $0.24\%$. The second effect is evaluated by varying the value of the number of neighbours ($k$ parameter) used in the kNN classifier when computing the weights, considering values between $30$ and $200$. 
This yields an uncertainty of $0.20\%$.

While asymmetries arising from regeneration and from mixing and \CP violation in the \mbox{$\Kz - \Kzb$} system do not affect the signal sample, they can affect the \decay{\Dz}{\KS\pip\pim} calibration sample. 
To evaluate the impact of the neglected \KS regeneration and oscillation, the dependence of $\ACP$ (after applying all weights) on the \KS flight distance is studied using the calibration data. No evidence of such a dependence is observed, and an upper limit of $0.05\%$ is set on the corresponding \ACP bias, which is negligible compared to the statistical and other systematic uncertainties.

An additional source of systematic uncertainty arises from the limited knowledge of the shape of the mass distributions. It is evaluated by fitting the simulated samples of pseudoexperiments with alternative models, whose fit quality is comparable to that of the baseline model.  The alternative models include: the sum of two Gaussian functions for the peaking component in the \Delm distribution; the sum of three Gaussian functions for the peaking component in the $m(\KS)$ distribution; and a polynomial multiplied by a threshold function for the nonpeaking component in the \Delm distribution. 
The largest observed variations are taken as the systematic uncertainty ranging from 0.1\% to 1.9\%, depending on the subsample. The systematic uncertainties due to the choice of fit model are assumed to be fully correlated across the data blocks, resulting in a total systematic uncertainty associated with the choice of the fit model of $0.27\%$.
A summary of all systematic effects can be found in Table~\ref{tab:syst_summary}.
\begin{table}[tb!]
    \centering
\caption{Summary of absolute systematic uncertainty sources on $\ACP(\decay{\Dz}{\KS\KS})$ in percent.}
    \label{tab:syst_summary}
    \begin{tabular}{l|c}
        Systematic source & Uncertainty [\%]\\
        \hline        
        Calibration sample size & 0.24 \\    
        Weighting procedure & 0.20 \\
        Fit model & 0.27 \\

    \end{tabular}
\end{table}

The data sample includes data blocks recorded with each polarity of the LHCb magnet. To check for potential uncompensated systematic effects, the asymmetries of the signal and calibration mode have been compared separately for the \MagUp and \MagDown magnet polarities. The results are found to be compatible within $1.5\sigma$.

\section{Results}
The measured $\ACP(\decay{\Dz}{\KS\KS})$ values for the different data blocks are shown in Table~\ref{tab:res}. All measurements are compatible with each other.
 \begin{table*}[tb!]
	\begin{center}
		\caption{Measurements of yields and $\ACP(\decay{\Dz}{\KS\KS})$ extracted from weighted fits to individual data blocks. The quoted uncertainty is statistical only.}
		\label{tab:res}
		\begin{tabular}{l|c c}
                Data block & Yield & \ACP [\%] \\
                \hline
                A & $1534 \pm 75$ & $\phantom{-}5.5 \pm 3.4$ \\
                B & $1639 \pm 56$ & $\phantom{-}0.8 \pm 3.2$ \\
                C & $1385 \pm 55$ & $-0.3 \pm 3.4$ \\
                D & $2915 \pm 85$ & $\phantom{-}0.3 \pm 2.4$ \\
                E & $3149 \pm 94$ & $\phantom{-}0.0 \pm 2.4$ \\
                F & $2544 \pm 77$ & $\phantom{-}4.6 \pm 2.6$ \\
                G & $1599 \pm 67$ & $\phantom{-}1.7 \pm 3.3$ \\
                H & $\phantom{0}911 \pm 54$ & $\phantom{-}5.6 \pm 4.3$ \\

		\end{tabular}
	\end{center}
\end{table*}
The weighted average is
$$\ACP(\KS\KS)= (1.86 \pm 1.04 \pm 0.41) \%,$$
where the first uncertainty is statistical, and the second is systematic, obtained by combining the individual sources of systematic uncertainty from Table~\ref{tab:syst_summary} as uncorrelated. 
This result is the most precise single measurement of this quantity to date, and it is compatible with \CP symmetry and with the average of previous determinations~\cite{CLEO:2000opx,LHCb-PAPER-2015-030,LHCb-PAPER-2020-047,CMS:2024hsv,Belle:2024vho,Belle-II:2025ezw}.
Any time-dependent effect can be neglected, as mentioned in Section~\ref{sec:Introduction}. It is anyway useful to provide the average decay time of the selected $\decay{\Dz}{\KS\KS}$ decays to allow the interpretation of this result in a wider global charm \CP violation scenario.
This value normalised to the \Dz lifetime~\cite{PDG2024}, is \mbox{$\langle t \rangle / \tau(\Dz) = 1.76 \pm 0.02$}. It is obtained by integrating all data blocks and purity bins, after performing sideband subtraction in the \Delm distribution. The contribution from secondary decays is subtracted, exploiting the same simulated \decay{\Dz}{\KS\KS} sample used for selections optimization. The quoted uncertainty includes statistical and systematic contributions.

Combining this result with the previous LHCb measurements~\cite{LHCb-PAPER-2015-030,LHCb-PAPER-2020-047}, assuming all systematic uncertainties are uncorrelated, the following value is obtained 
$$\ACP(\KS\KS)= (-0.37 \pm 0.78 \pm 0.29) \%.$$
The compatibility between the three measurements corresponds to a p-value of $1.2\%$.
Despite the similar integrated luminosity of the LHCb measurement based on the 2015–2018 dataset, and the exclusion of certain candidate categories included in the previous analysis~\cite{LHCb-PAPER-2020-047}, this result achieves better precision, thanks to the significantly improved efficiency of the upgraded trigger.

\section*{Acknowledgements}
%
%
\noindent We express our gratitude to our colleagues in the CERN
accelerator departments for the excellent performance of the LHC. We
thank the technical and administrative staff at the LHCb
institutes.
We acknowledge support from CERN and from the national agencies:
ARC (Australia);
CAPES, CNPq, FAPERJ and FINEP (Brazil); 
MOST and NSFC (China); 
CNRS/IN2P3 (France); 
BMFTR, DFG and MPG (Germany);
INFN (Italy); 
NWO (Netherlands); 
MNiSW and NCN (Poland); 
MCID/IFA (Romania); 
MICIU and AEI (Spain);
SNSF and SER (Switzerland); 
NASU (Ukraine); 
STFC (United Kingdom); 
DOE NP and NSF (USA).
We acknowledge the computing resources that are provided by ARDC (Australia), 
CBPF (Brazil),
CERN, 
IHEP and LZU (China),
IN2P3 (France), 
KIT and DESY (Germany), 
INFN (Italy), 
SURF (Netherlands),
Polish WLCG (Poland),
IFIN-HH (Romania), 
PIC (Spain), CSCS (Switzerland), 
and GridPP (United Kingdom).
We are indebted to the communities behind the multiple open-source
software packages on which we depend.
Individual groups or members have received support from
Key Research Program of Frontier Sciences of CAS, CAS PIFI, CAS CCEPP, 
Fundamental Research Funds for the Central Universities,  and Sci.\ \& Tech.\ Program of Guangzhou (China);
Minciencias (Colombia);
EPLANET, Marie Sk\l{}odowska-Curie Actions, ERC and NextGenerationEU (European Union);
A*MIDEX, ANR, IPhU and Labex P2IO, and R\'{e}gion Auvergne-Rh\^{o}ne-Alpes (France);
Alexander-von-Humboldt Foundation (Germany);
ICSC (Italy); 
Severo Ochoa and Mar\'ia de Maeztu Units of Excellence, GVA, XuntaGal, GENCAT, InTalent-Inditex and Prog.~Atracci\'on Talento CM (Spain);
SRC (Sweden);
the Leverhulme Trust, the Royal Society and UKRI (United Kingdom).

\clearpage

\clearpage

\addcontentsline{toc}{section}{References}
\bibliographystyle{LHCb}
\bibliography{main,standard,LHCb-PAPER,LHCb-CONF,LHCb-DP,LHCb-TDR}

\newpage
\centerline
{\large\bf LHCb collaboration}
\begin
{flushleft}
\small
R.~Aaij$^{38}$\lhcborcid{0000-0003-0533-1952},
A.S.W.~Abdelmotteleb$^{57}$\lhcborcid{0000-0001-7905-0542},
C.~Abellan~Beteta$^{51}$\lhcborcid{0009-0009-0869-6798},
F.~Abudin{\'e}n$^{57}$\lhcborcid{0000-0002-6737-3528},
T.~Ackernley$^{61}$\lhcborcid{0000-0002-5951-3498},
A. A. ~Adefisoye$^{69}$\lhcborcid{0000-0003-2448-1550},
B.~Adeva$^{47}$\lhcborcid{0000-0001-9756-3712},
M.~Adinolfi$^{55}$\lhcborcid{0000-0002-1326-1264},
P.~Adlarson$^{85}$\lhcborcid{0000-0001-6280-3851},
C.~Agapopoulou$^{14}$\lhcborcid{0000-0002-2368-0147}
% ...

{\footnotesize \it

% ...
$^{14}$Universit{\'e} Paris-Saclay, CNRS/IN2P3, IJCLab, Orsay, France\\
% ...
$^{38}$Nikhef National Institute for Subatomic Physics, Amsterdam, Netherlands\\
% ...
$^{47}$Instituto Galego de F{\'\i}sica de Altas Enerx{\'\i}as (IGFAE), Universidade de Santiago de Compostela, Santiago de Compostela, Spain\\
% ...
$^{51}$Physik-Institut, Universit{\"a}t Z{\"u}rich, Z{\"u}rich, Switzerland\\
% ...
$^{55}$H.H. Wills Physics Laboratory, University of Bristol, Bristol, United Kingdom\\
% ...
$^{57}$Department of Physics, University of Warwick, Coventry, United Kingdom\\
% ...
$^{60}$School of Physics and Astronomy, University of Glasgow, Glasgow, United Kingdom\\
$^{61}$Oliver Lodge Laboratory, University of Liverpool, Liverpool, United Kingdom\\
% ...
$^{69}$Syracuse University, Syracuse, NY, United States\\
% ...
$^{85}$Department of Physics and Astronomy, Uppsala University, Uppsala, Sweden, associated to $^{60}$\\
% ...
}
\end{flushleft}

\end{document}